\pdfoutput=1

\documentclass[11pt]{article}
\usepackage{authblk}
\usepackage[]{EMNLP2023}

\usepackage{times}
\usepackage{latexsym}

\usepackage[T1]{fontenc}

\usepackage[utf8]{inputenc}

\usepackage{microtype}

\usepackage{inconsolata}

\usepackage{tikz}

\usepackage{amsmath,amsfonts,bm}

\def\Figref#1{Figure~\ref{#1}}

\def\secref#1{section~\ref{#1}}

\def\eqref#1{equation~\ref{#1}}

\def\1{\bm{1}}

\def\ve{{\bm{e}}}

\def\vv{{\bm{v}}}

\def\mE{{\bm{E}}}

\def\mR{{\bm{R}}}

\DeclareMathAlphabet{\mathsfit}{\encodingdefault}{\sfdefault}{m}{sl}
\SetMathAlphabet{\mathsfit}{bold}{\encodingdefault}{\sfdefault}{bx}{n}

\def\sC{{\mathbb{C}}}

\def\sV{{\mathbb{V}}}

\usepackage[linesnumbered,ruled,vlined]{algorithm2e}
\usepackage{etoolbox}
\usepackage{xstring}
\usepackage{calc}
 
\newlength{\xalgowidth}
\newlength{\xalgoremainder}
\newlength{\xindentwidth}

\newenvironment{vAlgorithm*}[3][]{
  \setlength{\xalgowidth}{#2} 
  \setlength{\xindentwidth}{#3} 
  \setlength{\xalgoremainder}{\textwidth-\xalgowidth} 
  \SetCustomAlgoRuledWidth{\xalgowidth} 
  \IncMargin{\xindentwidth}
  \begin{algorithm*}[#1]
}
{
  \end{algorithm*} 
  \DecMargin{\xindentwidth}
}

\newenvironment{vAlgorithm}[3][]{
  \setlength{\xalgowidth}{#2} 
  \setlength{\xindentwidth}{#3} 
  \setlength{\xalgoremainder}{\columnwidth-\xalgowidth} 
  \SetCustomAlgoRuledWidth{\xalgowidth} 
  \IncMargin{\xindentwidth}
  \begin{algorithm}[#1] 
}
{
  \end{algorithm} 
  \DecMargin{\xindentwidth}
}

\makeatletter
\patchcmd{\@algocf@start}{%
\begin{lrbox}{\algocf@algobox}%
}{%
\rule{0.5\xalgoremainder}{\z@}
\begin{lrbox}{\algocf@algobox}%
\begin{minipage}{\xalgowidth}%
}{}{}
\patchcmd{\@algocf@finish}{%
\end{lrbox}%
}{%
\end{minipage}%
\end{lrbox}%
}{}{}
\makeatother
\usepackage{multirow}
\usepackage{hyperref}
\usepackage{url}
\usepackage{xcolor}
\usepackage{caption}
\usepackage{tcolorbox}
 \usepackage{enumitem}
 \usepackage{amsmath, amsfonts}
 \usepackage{listings}
\usepackage{graphicx}
\usepackage{setspace}
\SetKwInput{KwParam}{Hyperarameter}                
\SetKwInput{KwInput}{Input}                
\SetKwInput{KwOutput}{Output}              
\SetKwInput{KwInit}{Initialization}    
\SetKwComment{Comment}{/* }{ */}
\def\tabref#1{Table~\ref{tab:#1}}

\def\secref#1{Section~\ref{#1}}

\newcommand*{\affmark}[1][*]{\textsuperscript{#1}}
\newcommand{\cameraready}[1]{\textcolor{black}{#1}}
\newcommand{\approach}{\textsc{RNNS}\xspace}

\title{A Black-Box Attack on Code Models via Representation Nearest Neighbor Search}

\author[1\thanks{~~clark.zhang@huawei.com}]{Jie Zhang}
\author[2\thanks{\affmark[]\text{~~corresponding author: ma\_wei@ntu.edu.sg}}]{Wei Ma}
\author[3]{Qiang Hu}
\author[2]{Shangqing Liu}
\author[4]{Xiaofei Xie}
\author[3]{Yves Le Traon}
\author[2]{Yang Liu}
\affil[1]{Noah’s Ark Lab, Huawei}
\affil[2]{School of Computer Science and Engineering, Nanyang Technological University}
\affil[3]{The Interdisciplinary Centre for Security, Reliability and Trust, University of Luxembourg}
\affil[4]{School of Computing and Information Systems, Singapore Management University}

\begin{document}
\maketitle
\begin{abstract}
\cameraready{Existing methods for generating adversarial code examples face several challenges: limited availability of substitute variables, high verification costs for these substitutes, and the creation of adversarial samples with noticeable perturbations. To address these concerns, our proposed approach, \approach, uses a search seed based on historical attacks to find potential adversarial substitutes. Rather than directly using the discrete substitutes, they are mapped to a continuous vector space using a pre-trained variable name encoder. Based on the vector representation, \approach predicts and selects better substitutes for attacks. We evaluated the performance of \approach across six coding tasks encompassing three programming languages: Java, Python, and C. We employed three pre-trained code models~(CodeBERT, GraphCodeBERT, and CodeT5) that resulted in a cumulative of 18 victim models. The results demonstrate that RNNS outperforms baselines in terms of ASR and QT. Furthermore, the perturbation of adversarial examples introduced by RNNS is smaller compared to the baselines in terms of the number of replaced variables and the change in variable length. Lastly, our experiments indicate that RNNS is efficient in attacking defended models and can be employed for adversarial training.}

\end{abstract}

\maketitle
\section{Introduction}
\label{introduction}

Recently, since programming language can be seen as one kind of textual data and also inspired by the success of deep learning for text processing and understanding, researchers have tried to pre-train code models such as CodeBERT~\cite{feng2020codebert}, GraphCodeBERT~\cite{guo2020graphcodebert}, ContraBERT~\cite{liu2023contrabert} to help developers to solve multiple programming tasks, e.g., code search~\cite{gu2018deep, liu2023graphsearchnet}, code clone detection~\cite{white2016deep, li2017cclearner}, code summarization~\cite{ahmad2020transformer, liu2020retrieval}, and vulnerability detection~\cite{zhou2019devign}. Although these code models have achieved good performance on many code tasks, they are still suffering from robustness issues. A few adversarial attack methods have emerged to evaluate and improve the robustness of code models. 

There are certain considerations to be made. Firstly, code pre-training models are frequently deployed remotely, which limits access to the model parameters and renders white-box attacks infeasible. Secondly, among the numerous code-equivalent transformation methods, variable substitution exerts the most significant influence on the resilience of large code models while being the least detectable transformation~\cite{li2022closer}. As a result, black-box attack techniques based on variable substitution have emerged as a valuable avenue for research and multiple works have been proposed such as ALERT~\citep{naturalattack2022} and MHM~\citep{zhang2020generating}. 

However, these works have three limitations: 1) The number of substitute variables is limited and lacks diversity, which lowers the upper bound of the attack success rate. For example, ALERT employs 60 substitute variables for each variable, which are generated by a pre-trained model, and the substitute variables lack diversity. MHM also randomly selects 1500 words from a fixed dictionary as substitute variables. 2) The verification cost of substitute variables is high. To verify the attack effect of each substitute, it is necessary to replace the source variable with an adversarial sample and perform an actual attack on the victim model. ALERT uses a traversal method to select substitute variables, and in order to reduce the number of attacks, it limits the number of substitute variables; MHM uses a random sampling method to select substitute variables in order to reduce the number of attacks. Neither method is conducive to cost-effective attacks. 3) The generated adversarial samples have large perturbations. Each adversarial sample usually needs to replace multiple original variables to succeed in attacking, and MHM easily generates semantically incoherent and excessively long variable names.

To address the aforementioned challenges, in this paper, we propose a search-based black-box adversarial attack method to create challenging adversarial samples based on the search seed vector in the variable representation space, namely \textbf{Representation Nearest Neighbor Search} (\approach). Specifically, \approach, first utilizes publicly available real code datasets to construct a large original substitute set, denoted as $subs_{original}$. Then, based on the previous attack results, \approach predicts the search seed vector required for the next round of attacks and efficiently searches for the $k$ nearest substitutes to the seed vector from the large-scale original substitute set to form the $subs_{topk}$, where $k$ is much smaller than the size of the original substitute set. The generation process of the  $subs_{topk}$ does not involve attacking the victim model even once. \cameraready{Furthermore, the length and similarity of the substitute must adhere to specific perturbation constraints to prevent excessive deviations from $var$.}

To evaluate the effectiveness of \approach, we investigate three pre-trained code models, CodeBERT~\citep{feng2020codebert}, GraphCodeBERT~\citep{guo2020graphcodebert} and CodeT5~\citep{wang2021codet5}, and perform the attack on six code tasks in three programming languages, i.e., Java, Python, and C. The results on 18 victim models demonstrate that compared to the approaches MHM and ALERT, RNNS achieves a higher attack success rate (ASR) with a maximum of about 100\% improvement and 18/18 times as the winner. Meanwhile, \approach needs fewer query times~(QT) with 8/18 times as the winners. Furthermore, we analyze the quality of adversarial examples statistically and find that \approach introduces minor perturbations. In the end, we apply \approach to attack three defended models and find that our approach outperforms the baselines by up to 32.07\% ASR. We also use adversarial examples to improve the model's robustness through contrastive adversarial training.

\section{Preliminaries}
\label{sec:preliminaries}

\subsection{Textual Code Processing}
The nature of code data (in text format with discrete input space) makes it impossible to feed one code input $x$ directly into deep learning models. Thus, transferring code data to learnable continuous vectors is the first step in source code learning. \textbf{Dense encoding}~\cite{zhelezniak-etal-2020-estimating} is one common method used to vectorize textual code data. To do so, first, we need to learn a tokenizer that splits the code text into a token sequence which is called \textbf{Tokenization}. 
After tokenization, code $x$ is represented by a sequence of tokens, namely, $x=(s_0,...,s_j,..,s_l)$ where $s_i$ is one token. Then, the code vocabulary dictionary is built by using all the appeared tokens $s_i$, denoted $\sV$. After that, every word~(token) in $\sV$ is embedded by learned vectors $\vv_i$ with dimension $d$. Here, we use $\mE^{|\sV| \times d }$ to represent the embedding matrix for $\sV$. Finally, $x$ can be converted into a embedding matrix $\mR^{l \times d}=(\vv_0,...,\vv_j,..,\vv_l)$. After this code encoding, pre-trained code models based on the transformer take the matrix $\mR^{l \times d}$ as inputs and learn the contextual representation of $x$ for downstream tasks via pre-training such as Masked Language Modeling (MLM) and Causal Language Modeling (CLM).

\begin{figure}[]
\begin{center}
   \includegraphics[width=0.5\textwidth]{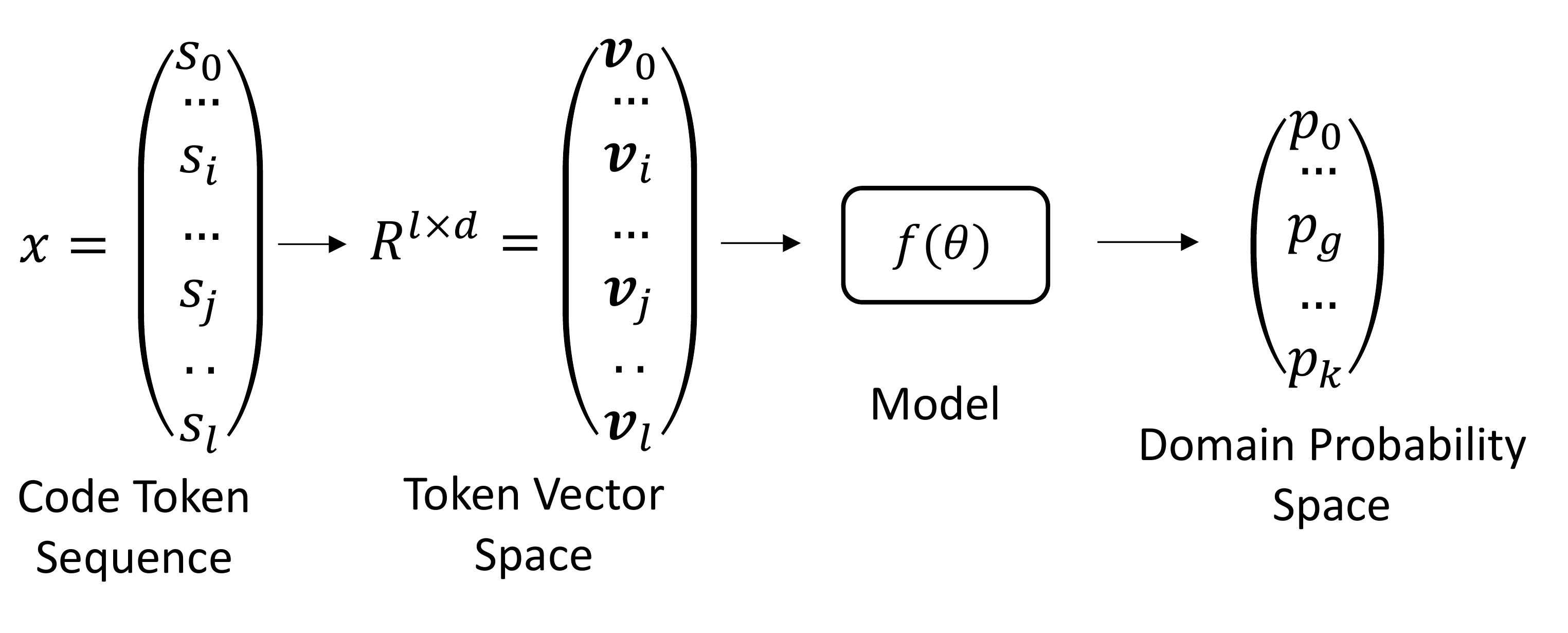} 
   \caption{One code model demo on the downstream task.}
   \label{fig:model_downstream}
\end{center}
\vspace{-2em}
\end{figure}

\Figref{fig:model_downstream} illustrates the main steps of the code processing models for the downstream classification tasks. First, we tokenize the textual code $x$ into a token sequence that is represented in a discrete integer space. Then, we map the discrete sequence ids into the token vector space $R^{l \times d}$. Next, we feed the token vectors into the task model $f(\theta)$. $f(\theta)$ is built on top of pre-trained models. Finally, we can predict the domain probabilities after fine-tuning.

\subsection{Problem Statement}
\label{sec:ps}

Since many critical code tasks are classification problems, e.g., defect prediction and code clone detection. In this paper, we focus on the adversarial attack for code classification tasks. Considering a code classification task, we use $f(x;\theta) \rightarrow y:\mR^{l \times d} \rightarrow \sC=\{i| 0 \le i \le n \}$ to denote the victim model that maps a code token sequence $x$ to a label $y$ from a label set $\sC$ with size $n$, where $l$ is the sequence length and $d$ is the token vector dimension, and $i$ is one integer. By querying dictionary dense embedding $\mE^{|\sV \times d|}$, a code token sequence $x=(s_0,...,s_j,..,s_l)$, is vectorized into $\mR^{l \times d}$. Adversarial attacks for code models create an adversarial example $x'$ by modifying some vulnerable tokens of $x$ with a limited maximum perturbation $\epsilon$ to change the correct label $y$ to a wrong label $y'$. Simply,  we get a perturbed $x'$ by modifying some tokens in $(s_0,...,s_j,..,s_l)$ such that $f(x';\theta) \ne f(x;\theta)$ where $x'=x+\sigma$ and $x'$ has to have the same behavior with $x$, $+$ represent perturbation execution, $\sigma$ is the perturbation code transformation for  $(s_0,...,s_j,..,s_l)$, and $\sigma \le \epsilon$. We target the more practical attacking scenario -- black-box attack that requires less information. We assume we cannot access the model parameters and can only utilize the final output of model $f(x;\theta)$ to conduct the attack.

\section{Methodology}
\label{methodology}

\subsection{\textbf{Motivation}}
As mentioned in the introduction, the current methods face three limitations: 1) there is a limited number of substitute variables; 2) there is a high verification cost associated with substitute variables; and 3) the generated adversarial samples often exhibit large perturbations. Among these limitations, the second one holds the utmost significance as it significantly impacts both the first and third limitations.
Due to the high cost involved, it becomes challenging to generate diverse adversarial examples within a reasonable budget. Additionally, attackers tend to introduce large perturbations without employing any perturbation constraints in order to maximize their attacks. 

To address these limitations, the \textbf{first} question arises: "\textit{Could we substantially reduce the verification cost while allowing for unrestricted diversity of substitute variables and minimizing perturbations?}"
To delve into the reasons behind the second limitation, we need to analyze its underlying factors. The low verification efficiency of the substitute set stems from the fact that each substitute can only be verified by constructing an adversarial sample to replace the original variable and then launching an actual attack on the victim model. This realization leads to the \textbf{second} question: "\textit{Is it feasible to predict the attack effect of a substitute instead of constructing an adversarial sample to attack the victim model?}"

Given input code $x$ and one of its variables $var$, different substitutes can be used to replace it to obtain different adversarial samples. After attacking the victim model, the probability of the label will also change. Conversely, if we want to reduce the probability of this label, the \textbf{third} question is following, "\textit{how to choose relatively better substitutes that can reduce the model confidence from a large-scale original substitute set?}". It is possible to select good substitutes without actual attack if we can forecast, which is implemented by \approach.

The core idea of \approach is maintaining a search seed updated based on the historical attack. The search seed is employed to search next adversarial substitutes that are possible to attack successfully. Since substitutes are discrete and cannot be directly involved in calculations, we first use a variable name \cameraready{pre-trained} encoder denoted as $E$ to map substitutes to a unified continuous representation vector space. Then, based on the representation vectors of substitutes that have participated in the attack, we predict the search seed vector $e_{seed}$ for the next round of the substitute selection. Finally, we calculate the similarity between $e_{seed}$ and the representation vector of substitutes and then select relatively better substitutes. For specific details, please refer to \secref{subsec:pss}.

\subsection{\textbf{Representation Nearest Neighbor Search}} 

\begin{vAlgorithm}[t!]{0.45\textwidth}{1em}
\setstretch{0.15}
\tiny
\SetAlCapFnt{\tiny}
\caption{\tiny \approach}
\label{alg:findAdv}
\DontPrintSemicolon
\SetAlgoLined
\SetNoFillComment
  \KwParam{ maximum attacking iteration $max\_itr$}
  \KwInput{input code $x$ with ground label $y$,  original substitute set $subs_{original}$}
  \KwOutput{adversarial example $x'$, attacking result $is\_suc$}
  $x' = x$  \\
  $prob_{min}=1.0$\\
  {$vars  = ExtractVar(x)$} \\
  {$vars = \scalebox{0.9}{RankVarsWithUncertainty}(vars)$}\\
 \For{$var \in vars$ }
 { 
  $sub_{pre}=var$\\
  $sub_{cur}=var$\\
  $\Delta \ve_{smo} = 0 $\\
  $i=0$ \\
  $is\_suc=false$\\
  \While{$i < max\_itr$}
    {
    \scalebox{0.9}{$\ve_{seed}, \Delta \ve_{smo}= PredictSeed(sub_{pre}, sub_{cur},\Delta \ve_{smo})$ }\\
    \scalebox{0.9}{$subs_{topk}=SearchTopkSub(\ve_{seed},subs_{original},var)$}\\
    \For{$sub \in subs_{topk}$}
    {
    $x'_{tmp}=Replace(x',sub_{cur},sub)$ \\
    $prob_y, y'=f(x'_{tmp};\theta)$ \\
    \If{$prob_y<prob_{min}$} 
    {
        {$x'=x'_{tmp}$} \\
        {$sub_{pre}=sub_{cur}$} \\
        {$sub_{cur}=sub$} \\
        {$prob_{min}=prob_y$} \\      
    }
    \If{$y!=y'$}
    {   
        {$is\_suc=true$} \\
        return $x'$, $is\_suc$ \\
        
    }
    }
    }
 }
return $x'$, $is\_suc$
\end {vAlgorithm}

Algorithm \ref{alg:findAdv} shows the workflow of our approach, First, we collect the original substitute set from public real code, following the process described in \secref{subsec:closs}. We extract variables from the input code and sort them according to their uncertainty, referring to \secref{subsec:cu}~(Line 3-4). We replace variables in sequence to form attack samples (Line 5). For a given $var$, we first initialize the optimal substitute for this current iteration $sub_{cur}$ and the optimal substitute for the previous iteration $sub_{pre}$ to the $var$. Then, we initialize the accumulated smooth increment of the representation vector $\Delta \ve_{smo}$ to a zero vector. $\Delta \ve_{smo}$ is used to record the historical representation change of the search seed $\ve_{seed}$.  We now commence the iterative attack process, as delineated in Line 11. We predict the search seed vector $\ve_{seed}$ with the process described in \secref{subsec:pss}~(Line 12), and then extract $topk$ substitutes based on $\ve_{seed}$ to form the candidate substitutes $subs_{topk}$ with the process described in \secref{subsec:sts}~(Line 13). Subsequently, we replace $sub_{cur}$ in $x'$ with each substitute in $subs_{topk}$ to obtain the corresponding temporary adversarial sample $x'_{tmp}$~(Line 14-15). $x'$ is the current code that we are trying to attack and it is initialized with the original code $x$. We use $x'_{tmp}$ to attack the victim model and obtain the probability $prob_y$ of the ground-truth label y and predicted label $y'$~(Line 16). If the probability of the ground-truth label y hits a new low ($<prob_{min}$), we update $x'$, $sub_{pre}$, $sub_{cur}$ and $prob_{min}$~(Line 17-22). $prob_{min}$ records the minimum probability of label $y$ during the attack process. If $x'_{tmp}$ causes the victim model to predict an incorrect label, this attack is successful and returns the successful adversarial sample~(Line 23-26); otherwise, proceed to the next iteration until all variables have completed iteration and return the final adversarial sample and attack result~(Line 30).

\subsubsection{Collecting Large Original Substitute Set}
\label{subsec:closs}
\cameraready{
We have developed a tool for variable extraction that leverages the tree-sitter framework\footnote{{\scriptsize \url{https://tree-sitter.github.io/tree-sitter}}}. This tool, henceforth denoted as $ExtractVar$ (see Line 3), operates in three distinct steps. In the first step, we extract all variables from the current dataset and then filter out duplicates. During the second step, each valid variable is tokenized, and we compute the embedding for each token using the variable-name encoder $E$ that is pre-trained on CodeSearchNet\footnote{{\scriptsize \url{https://huggingface.co/datasets/code_search_net}}}. We then apply a mean pooling operation on these tokens to determine the variable's embedding. In the third step, we retain all the chosen variables along with their associated embeddings as the initial substitute set, represented as $subs_{original}$.}

\subsubsection{Computing  Uncertainty}
\label{subsec:cu}
Given a specific code $x$, we replace each instance of $var \in x$ with \cameraready{ a set of predefined fixed variables $VarArray$}, resulting in a set of mutated codes denoted as \cameraready {$X^{mutated}_{var}$}. These mutated codes are subsequently utilized to query the victim model, allowing us to obtain the probability distribution for each class. A greater variance in the distribution signifies increased uncertainty for $var$, suggesting that $var$ should be prioritized for replacement. The uncertainty associated with $var$ is defined as follows: \cameraready{$$uncertainty_{var}=\frac{1}{C}\sum_{i=1}^C variance(P^{i}_{var})$$, where $P^{i}_{var}=\{p^{i}_{var}(x)| \forall x \in X^{mutated}_{var}\}$, $C$ is the number of labels, $p^i_{var}(x)$ is the model probability for label $i$ given the mutated code $x$, and $variance$ denotes the standard variance function. A larger and more diverse $X^{mutated}_{var}$ ensures a closer approximation of $uncertainty_{var}$ to the true value.} It is important to note, however, that the magnitude of the change length must not be excessively large, as this would result in all probability changes converging to a single point. This is because samples subjected to large changes deviate significantly from the original, leading to a substantial decrease in the model confidence levels. Subsequently, we arrange the variables in descending order based on their uncertainties. The greater the uncertainty of a variable, the more valuable it is for attack. This process is denoted as $RankVarsWithUncertainty$ at line 4. 
\cameraready{In our implementation, the size of this variable array $VarArray$ is 16, and the variable length ranges from 1 to 5.
}

\subsubsection{Predicting Search Seed}
\label{subsec:pss}
To filter out superior substitutes from the substantial $subs_{original}$, it becomes necessary to predict the search seed within the substitute representation vector space. Given the optimal substitute $sub_{cur}$ of the current round, the optimal substitute $sub_{pre}$ from the previous round, and the accumulated smooth increment of the representation vector, denoted as $\Delta \ve_{smo}$, from all preceding rounds of iteration, we initially compute the increment of the representation vector in the current round, $\Delta \ve$:
\[\Delta \ve=E(sub_{cur}) - E(sub_{pre})\], where $E$ is variable name encoder, trained \cameraready{on CodeSearchNet} by masked language modelling independently so that \approach is independent of victim downstream-task models. Then we update the $\Delta \ve_{smo}$,  \[\Delta \ve_{smo} = (1-\alpha) \Delta \ve_{smo} + \alpha \Delta \ve \], where $\alpha$ is a smooth rate limited 0 to 1,
Finally, we predict the search seed $\ve_{seed}$: \[\ve_{seed}=E(sub_{cur}) + \Delta \ve_{smo}\]
This process is denoted as $PredictSeed$ at line 12.

\subsubsection{Searching Top-K Substitutes}
\label{subsec:sts}
Initially, we filter out substitutes from $subs_{original}$ that comply with two constraints: 1) $1-sim(E(sub), E(var))<\epsilon$ and 2) $|len(sub) - len(var)| < \delta$, where $var$ refers to the original variable in the input code that is to be replaced, $sim(.)$ is the similarity calculation function. $E(.)$ is the variable name encoder, and $len(.)$ is used to calculate the length of the variable name. Then, we calculate the similarity between the search seed $e_{seed}$ and the substitutes that are filtered by the two constraints and select the $k$ most similar substitutes to form $subs_{topk}$. This process is denoted as $SearchTopkSub$ at line 13. \cameraready{In our experiment, $\epsilon=0.15$, $\delta=4$, $k=60$, $sim(.)$ is cosine similarity.}

\section{Experimental Setup}
\label{experiments}

\begin{table}[!t] \large
\begin{center}
\scalebox{0.5}{
\begin{tabular}{l|c|c|c|c}
\hline
     \bf Task  & \bf Train / Val /  Test & \bf CodeBERT & \bf GraphCodeBERT &\bf CodeT5 \\ \hline
 \bf Defect &   21,854 / 2,732 / 2,732  &    63.76   &  63.65 &    67.02        \\ 
 \bf Clone         &   90,102 / 4,000 / 4,000  &   96.97 &  97.36  &    97.84          \\ 
\bf Authorship    &   528 / – / 132      & 82.57 & 77.27 & 88.63          \\ 
\bf C1000         &   320,000 / 80,000 / 100,000   &    82.53   &   83.79  &  84.46          \\ 
\bf Python800     &   153,600 / 38,400 / 48,000    &    96.39  &    96.29  &  96.79          \\ 
\bf Java250       &   48,000 / 11,909 / 15,000     &   96.91   &   97.27   &   97.72          \\ \hline
\end{tabular}
}
\end{center}
\caption{Datasets and Victim Model Performance (Accuracy, \%).}
\label{tab:datasetmodel}
\end{table}

\textbf{Dataset and Model.} To study the effectiveness and efficiency of RNNS, we conduct experiments on three popular programming languages~(C, Python, and Java). For the datasets, we employed six widely studied open-source datasets that cover four important code tasks. Specifically, BigCloneBench~\citep{wang2020detecting} is one code clone detection dataset named Clone. Devign~\citep{zhou2019devign} is a dataset used for vulnerability detection, named Defect. For authorship prediction, we use the dataset provided by ~\citep{10.1007/978-3-319-66402-6_6}. Besides, we utilize three problem-solving classification tasks, Java250, Python800, and C1000, provided by ProjectCodeNet~\citep{puri2021codenet}. For all the datasets (except for authorship prediction which does not have enough data samples), we follow the original papers to split the data into the training set, validation set, and test set. Authorship prediction only has two split parts, training data and test data.

For the code models, we follow the previous work~\cite{naturalattack2022} and investigate two pre-trained models CodeBERT~\citep{feng2020codebert}, and GraphCodeBERT~\citep{guo2020graphcodebert}. Besides, we add one more powerful model CodeT5~\citep{wang2021codet5} in our study. Table~\ref{tab:datasetmodel} summarizes the details of our employed datasets and fine-tuned models. 

\noindent\textbf{Evaluation Metric.} To evaluate the effectiveness of adversarial attack methods, we employ the commonly used attack success rate (ASR)~\cite{naturalattack2022} as the measurement. To evaluate the efficiency of the attack methods, we use query times (QT) to check the average number of querying the victim model for one input code. Finally, we use the change of replaced-variable length and the number of replaced variables to study the quality/perturbation of adversarial examples. A smaller score means the attack method can generate adversarial examples with less perturbation injection.

\noindent\textbf{Baseline.} We compare RNNS with two black-box attack baselines, MHM~\citep{zhang2020generating} and NaturalAttack~(ALERT)~\citep{naturalattack2022}. MHM is a sampling search-based black-box attack that generates the substitutes from the vocabulary based on lexical rules for identifiers. MHM employs synthesized tokens as the candidates of substitutes, which could introduce meaningless variable names. ALERT is a recently proposed attack method that combines greedy attack and genetic algorithm to find the substitutes. \cameraready{We also use two textual attack algorithms PSO~\citep{zang2020word} and LSH~\citep{maheshwary2021strong} as minor baselines, since they are not designed for code models. }

\noindent \textbf{Implementation.} We implement our approach in PyTorch and run all experiments on 32G-v100 GPUs. We reuse the source code from the baselines.  \cameraready{We make our implementation ~\footnote{{\scriptsize \url{https://github.com/18682922316/RNNS-for-code-attack}}} publicly available.}

\section{Results Analysis}
\label{sec:evaluation}

\begin{table*}[] 
\centering
\scalebox{0.7}{
\begin{tabular}{l|cc|cc|cc}
\hline
\multirow{2}{*}{Task+Model}     & \multicolumn{2}{c|}{\bf ALERT}   & \multicolumn{2}{c|}{\bf MHM}    & \multicolumn{2}{c}{\bf \approach}   \\ 
          & ASR & QT & ASR & QT  & ASR & QT \\ \hline
Clone+CodeBert      & 28.67 &	2155.39  & 39.66 & 972.15  &  \bf 46.50   & \bf 666.48  \\ 
Clone+GraphCodeBert     &  10.40 &	1466.68  &  \underline{9.58}   & \bf 490.99  & \bf 41.28  &  1122.01  \\ 
Clone+CodeT5             &  29.20 & 2359.70  &  38.79  & 1069.06 &  \bf 39.61  & \bf 895.79    \\ 
Defect+CodeBert          &  52.29 &	1079.68  &  50.51  & 862.18  & \bf 69.18  & \bf 588.35  \\ 
Defect+GraphCodeBert     &  74.29 & 621.77 &  75.19  & 539.93  & \bf 81.63  & \bf 404.73  \\ 
Defect+CodeT5            &  76.66 & 721.02  &  86.51  & \bf 344.08  & \bf 89.45  & 344.29     \\ 
Authorship+CodeBert      &  34.98 & \bf 682.57 &  64.70   & 775.11   & \bf 73.39  & 1029.59     \\ 
Authorship+GraphCodeBert  & 58.82 & 1227.36  &  75.49  & \bf 632.10   & \bf 80.39  & 696.64     \\ 
Authorship+CodeT5         & 64.95 & 1078.40 &  66.97  & \bf 715.89   & \bf 71.79  &    970.44   \\ 
Java250+CodeBert         &  50.50 & 958.96  &  74.03  & 961.60  & \bf 75.12  & \bf 815.91   \\ 
Java250+GraphCodeBert    &  46.74 & 1026.15  &  46.05  & 946.52  & \bf 72.30   & \bf 853.74    \\ 
Java250+CodeT5           &  52.04 & 1189.42  &  30.59  & 1107.95 & \bf 63.80   & \bf 1049.46  \\ 
Python800+CodeBert       &  58.30 & \bf 513.63  &  56.67  & 919.37  & \bf 77.88  & 514.19    \\ 
Python800+GraphCodeBert  &  51.87 & \bf 577.70 &  54.15  & 917.92  & \bf 71.42 &   730.14    \\ 
Python800+CodeT5         &  52.84 & 777.20  &  36.95  & 1127.44 & \bf 69.07  & \bf 662.28    \\ 
C1000+CodeBert           &  53.50 & 525.43  &  59.75  & \bf 340.88  & \bf 72.96  & 537.76       \\ 
C1000+GraphCodeBert       & 52.68 & \bf 566.18 &  45.93   & 837.09    & \bf 72.23    &	634.27          \\ 
C1000+CodeT5              & 47.86 & 843.33  &  36.45  &\bf 668.15  & \bf 59.00     & 697.06          \\ \hline
 Count                   & 0/18 & 4/18 &    0/18    &   6/18    &  18/18      &    8/18        \\ \hline
\end{tabular}
}
\caption{Comparison results with MHM, and ALERT, ASR \%. Count: the number of best results achieved.}
\label{tab:baseline_comparision}
\vspace{-4mm}
\end{table*}

\subsection{Attack Effectiveness and Efficiency}
We compare RNNS with two methods MHM~\cite{zhang2020generating} and NaturalAttack (ALERT)~\cite{naturalattack2022} on six datasets and 18 victim models that have been fine-tuned for the downstream tasks. Table~\ref{tab:baseline_comparision} shows the comparison results where the last row \textit{Count} indicates how many times this method achieves the best results. We can see that \approach achieves the best performance for 18/18 times in terms of ASR, and the lowest cost for 8/18 times in terms of QT in Table~\ref{tab:baseline_comparision}. Both of the indicators are better than the baselines. The two baselines have zero best ASR for all victim models and all datasets. The lowest QTs achieved by ALERT and MHM are 4 and 6, respectively. We conclude that for effectiveness and efficiency, \approach outperforms ALERT and MHM in all cases. Especially, MHM and ALERT fail to attack GraphCodeBERT on BigClone dataset, and only have  $9.58\%$ and $10.4\%$ ASR respectively, while \approach has more than $40\%$ ASR. \approach has almost two times larger ASR than MHM on Java250+CodeT5 and Python800+CodeT5. 

\cameraready{It should be noted that high ASR is not due to large QT. As shown in Table~\ref{tab:baseline_comparision}, the three groups of experiments with the most QTs are Clone+GraphCodeBert, Java250+CodeT5, and Authorship+CodeBert, with ASRs of 41.28\%, 63.80\%, and 73.39\%, respectively, which are not the highest. On the contrary, Defect+CodeT5 has the highest ASR of 89.45\%, but QT is the smallest. Therefore, there is no absolute causal relationship between QT and ASR.}

\subsection{Perturbation of Adversarial Example}
We conduct a study about the quality of the adversarial examples to check if RNNS can generate looking-normal code, e.g., avoiding naively increasing the variable name length. To do so, firstly, we count the average length of the original variable and adversarial variables as demonstrated by \tabref{variable_len}. We also compute the mean and variances of their difference. Besides, we compute the average number of the replaced variables for the successful attack as shown in \tabref{variable_num}. Low values mean the inputs are modified less, and high qualities. 

\begin{table*}[!t]
\centering
\scalebox{0.65}{
\begin{tabular}{l|ccc|ccc|ccc}
\hline
\multirow{2}{*}{Task+Model}        & \multicolumn{3}{c|}{\bf \approach}      & \multicolumn{3}{c|}{\bf MHM}       & \multicolumn{3}{c}{\bf ALERT}  \\ 
                         &  \small{Var Len} & \small{Adv Var Len} & \small{Difference} &  \small{Var Len} & \small{Adv Var Len} & \small{Difference} & \small{Var Len} & \small{Adv Var Len} & \small{Difference} \\ \hline
Clone+CodeBert           & 6.12          & 6.79     & 2.35 $\pm$ 4.50          & \bf 6.47          & \bf 10.6     & 6.34   $\pm$ 10.98   & 5.91 &	6.21 &	1.32 $\pm$	2.02    \\ 
Clone+GraphCodeBert      & 6.32          & 6.97     & 2.54 $\pm$ 6.43         & \bf 6.58          & \bf 10.41    & 6.82   $\pm$ 21.67    & 5.50 &	5.93 &	1.45 $\pm$	2.49    \\ 
Clone+CodeT5             & 6.45          & 6.69     & 2.51 $\pm$ 8.30          & \bf 6.46          & \bf 10.46    & 6.17   $\pm$ 25.78    & 6.25	 & 6.61	 & 1.32 $\pm$	2.72    \\ 
Defect+CodeBert          & 4.64          & 5.44     & 2.08 $\pm$ 2.49         & 4.44          & \bf 9.59     & 6.57   $\pm$ 28.78    & \bf 4.85  &	5.06  &	1.36 $\pm$	1.93    \\ 
Defect+GraphCodeBert     & 4.08          & 5.34     & 2.13 $\pm$ 1.83         & 4.37          & \bf 9.73     & 6.48   $\pm$ 26.51   &  \bf 4.47	 & 5.22  &	1.33 $\pm$	1.83    \\ 
Defect+CodeT5            & 3.95          & 5.17     & 2.03 $\pm$ 1.93         & 4.33          & \bf 9.81     & 6.59   $\pm$ 29.98     & \bf 4.36 &	5.01 &	1.27 $\pm$	1.57    \\ 
Authorship+CodeBert      & 3.81          & 5.18     & 2.28 $\pm$ 1.56         & 3.97          & \bf 7.94     & 5.45   $\pm$ 16.72    & \bf 4.42	 & 5.35	  & 1.40 $\pm$	    2.25    \\ 
Authorship+GraphCodeBert & 3.69          & 5.23     & 2.36 $\pm$ 1.71         & \bf 4.39          & \bf 7.64     & 5.24    $\pm$ 15.38    & 3.74  &	4.46  &	1.22 $\pm$	1.82     \\ 
Authorship+CodeT5        & \bf 3.95          & 5.18     & 2.03  $\pm$ 2.66         & \bf 3.95          & \bf 7.98     & 5.59  $\pm$ 20.94     & 3.81  &	4.50	 & 1.22 $\pm$	1.62    \\ 
Java250+CodeBert         & 2.35          & 4.22     & 2.11  $\pm$ 1.02         & 3.21          & \bf 6.50      & 4.34  $\pm$ 15.20     &  \bf 3.22  &	3.65  &	0.94 $\pm$	1.63    \\ 
Java250+GraphCodeBert    & 2.48          & 4.31     & 2.13  $\pm$ 1.07         & \bf 3.13          & \bf 6.59     & 4.42  $\pm$ 14.84      & 3.05  &	3.50	 & 0.98 $\pm$	1.54    \\ 
Java250+CodeT5           & 2.76          & 4.47     & 2.10   $\pm$ 1.17         & \bf 3.20           & 6.54     & 4.33   $\pm$ 14.60      & 3.16  &	\bf  7.31  &	4.41 $\pm$	18.73     \\ 
Python800+CodeBert       & 1.50           & 3.54     & 2.21  $\pm$ 1.02         & \bf  1.97          & \bf 5.11     & 3.64   $\pm$ 9.06      & 1.78	  &  2.27  &	0.64 $\pm$	1.34    \\ 
Python800+GraphCodeBert  & 1.88          & 3.90      & 2.18  $\pm$ 0.78         & \bf  1.99          & \bf  6.01     & 4.46   $\pm$ 16.52     & 1.80  &	2.33  &	0.76 $\pm$	1.30    \\ 
Python800+CodeT5         & 1.65          & 3.59     & 2.13  $\pm$ 0.95         & \bf 1.97          & 4.95     & 3.49   $\pm$ 8.18       & 1.88  &	\bf 5.84  &	4.10 $\pm$	12.64   \\ 
C1000+CodeBert           & 1.58          & 3.44     & 2.08  $\pm$ 0.88         & \bf 2.41          & \bf 5.05     & 3.65  $\pm$ 12.02      &	2.13  &	2.52  &	0.67 $\pm$	1.17    \\ 
C1000+GraphCodeBert      & 1.60           & 3.59     & 2.10   $\pm$0.85         & \bf 2.39          & \bf 5.35     & 3.90    $\pm$ 12.98      & 2.18  &	2.67  &	0.66 $\pm$	1.23   \\ 
C1000+CodeBert           & 1.38          & 3.33     & 2.02   $\pm$ 0.85         & \bf 2.36          & 4.82     & 3.39  $\pm$ 10.98      &	2.10	& \bf 6.56  &	4.74 $\pm$	13.24    \\ \hline
\end{tabular}
}
\caption{Replaced-variable length comparison, $mean \pm variance$.}
\label{tab:variable_len}
\vspace{-4mm}
\end{table*}

In Table~\ref{tab:variable_len}, the 2nd, 5th, and 8th columns are the average length for original variables (named \textit{Var Len}) that are replaced. The 3rd, 6th, and 9th columns are the average lengths for adversarial variables (named \textit{Adv Var Len}). The 4th, 7th, and 10th columns are the average and variance ($ mean \pm  variance$) of the absolute length difference between original variables and adversarial variables (named \textit{Difference}). We observe that MHM prefers to replace the long-length variables while \approach likes replacing short-length variables if we compare the 2nd and 5th columns. Meanwhile, the change of variable length from \approach is less than MHM. MHM introduces the average length difference of 3.39-6.82 while \approach only has 2.02-2.54. MHM has much higher variances than \approach in the length change. ALERT uses shorter adversarial variable names than \approach with less change because it uses the pre-trained model to generate the replacements that are close to the replaced variables. 

Table~\ref{tab:variable_num} statistically shows the number of replaced variables. It can be seen that \approach replaces around an average of 3.6 variables with a smaller variance of around (3.4-4.6) while MHM needs to modify about an average of 5.4 variables with a larger variance ($\ge$ 11.14). ALERT also replaces more variables to attack models than \approach and MHM. 
\approach introduces less or equal perturbation than the baselines in terms of length change and change number.

\begin{table*}[!t]
\centering
\scalebox{0.65}{
\begin{tabular}{l|lll|lll|lll}
\hline
\multirow{3}{*}{Task} & \multicolumn{3}{c|}{\bf CodeBERT}     & \multicolumn{3}{c|}{\bf GraphCodeBERT}   & \multicolumn{3}{c}{\bf CodeT5}  \\ 
         & \approach   & MHM  & ALERT & \approach  & MHM & ALERT & \approach  & MHM & ALERT  \\ \hline
Clone           & 3.55  $\pm$ 4.60   &    6.72 $\pm$  16.57 & 6.86 $\pm$ 18.85 & 4.12   $\pm$  4.94  & 6.21 $\pm$   15.13 & 6.95	 $\pm$ 18.99 & 3.43  $\pm$  5.00   &5.68  $\pm$   14.01 & 7.65 $\pm$ 	25.57\\
Defect          &3.39  $\pm$ 4.96  &    2.78 $\pm$  7.89  & 3.49 $\pm$	3.99 & 2.67   $\pm$  1.75  & 2.84 $\pm$   9.50   & 4.10 $\pm$ 	11.05 &2.51  $\pm$  1.45  &2.16  $\pm$  3.58 & 3.49	 $\pm$ 3.99 \\
Authorship      &4.24  $\pm$ 7.47  &    7.52 $\pm$  25.82  & 6.60 $\pm$	22.96 & 3.65  $\pm$  3.32  & 6.67 $\pm$   22.29 & 7.75 $\pm$ 	33.12 & 4.39  $\pm$  9.00  &5.72  $\pm$  13.02 & 	6.06 $\pm$ 	18.74\\
Java250         &3.87  $\pm$ 4.70   &    7.11 $\pm$  21.18 & 7.82 $\pm$	28.96& 3.87   $\pm$  4.25  & 6.41 $\pm$   16.24 & 7.83 $\pm$ 	25.06 & 4.71  $\pm$  6.87  &7.04  $\pm$ 15.29 & 	8.92 $\pm$ 	25.97 \\ 
Python800       &3.06  $\pm$ 1.87  &    5.21 $\pm$  12.28 & 4.96 $\pm$ 8.47 & 4.12   $\pm$  3.68  & 5.00 $\pm$   10.83 & 4.63 $\pm$ 	6.76 & 3.57   $\pm$  3.04  &5.29  $\pm$ 13.51 & 	6.18 $\pm$ 	11.45\\
C1000           &3.00     $\pm$ 1.86  &    4.42 $\pm$  7.49 & 4.13 $\pm$ 5.59 & 3.37   $\pm$  2.38  & 5.14   $\pm$   7.30  & 	4.88 $\pm$ 	6.24 & 3.39  $\pm$  2.48  & 5.20  $\pm$  7.43    &  5.43	 $\pm$ 6.99  \\ \hline
mean & \bf 3.52 $\pm$	4.24 & 5.63 $\pm$ 15.21 & 5.65 $\pm$ 14.80\ & \bf 3.63 $\pm$	3.39 & 5.38  $\pm$	13.55 & 6.02 $\pm$ 16.87 & \bf 3.67 $\pm$	4.64 & 5.18 $\pm$	11.14 & 6.29 $\pm$ 15.45 \\ \hline
\end{tabular}
}
\caption{ Replaced-variable number comparison, $mean \pm variance$ }
\label{tab:variable_num}
\end{table*}

\begin{figure*}[!t]
\begin{center}
   \includegraphics[width=\textwidth]{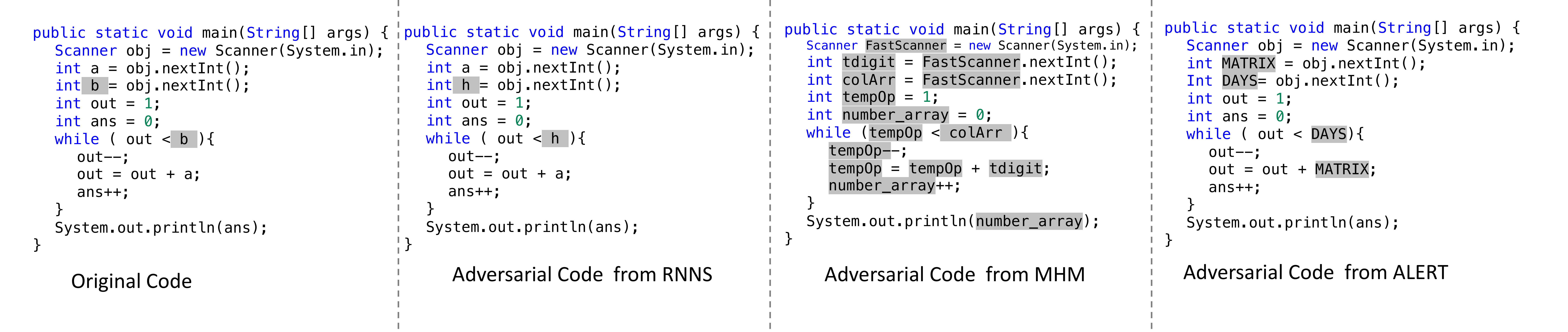} 
   \caption{Case study. Original vs. RNNS vs. MHM vs. ALERT}
   \label{fig:example_adv}
\end{center}
\end{figure*}

\Figref{fig:example_adv} shows one example of RNNS, MHM, and ALERT attack successfully from the Java250 dataset. The changes are highlighted by shadow markers. RNNS only renames one variable \textbf{b} to \textbf{h}, ALERT renames two variables, while MHM almost renames all variables and also prefers longer names.

\subsection{Ablation Study}
We remove the two search constraints in \secref{subsec:sts}, denoted this variant of \approach as \approach-Unlimited. Table~\ref{tab:ablation} shows the comparing results between \approach-Unlimited and \approach. \approach-Unlimited gets the first place for all the tasks in terms of ASR. ASR can be improved by a maximum of 8.35\% and a minimum of about 2\% after removing limitations. For QT, \approach-Unlimited only loses 3 times among 18 evaluations. The improvement of \approach-Unlimited is not surprising with respect to ASR and QT. Because \approach-Unlimited can search the adversarial examples in the non-similar real names and use very long variable names. 

\begin{table*}[!t]
\centering
\scalebox{0.7}{
\begin{tabular}{l|cccc|cccc|cccc}
\hline
\multirow{3}{*}{Task} & \multicolumn{4}{c|}{\bf CodeBERT}                                                                   & \multicolumn{4}{c|}{\bf GraphCodeBERT}                                                               & \multicolumn{4}{c}{\bf CodeT5}                                                                      \\ 
                  & \multicolumn{2}{c}{RNNS-Unlimited}                       & \multicolumn{2}{c|}{RNNS}    & \multicolumn{2}{c}{RNNS-Unlimited}                       & \multicolumn{2}{c|}{RNNS}     & \multicolumn{2}{c}{RNNS-Unlimited}                       & \multicolumn{2}{c}{RNNS}     \\ 
                  & ASR   & QT      & ASR   & QT     & ASR   & QT      & ASR   & QT      & ASR   & QT      & ASR   & QT      \\ \hline
Defect            & \bf 72.29 & 590.98  & 69.18 & \bf 588.35 & \bf 87.77 &  \bf 381.82  & 81.63 & 404.73  &  \bf 91.64 &	\bf 338.41	 &  89.45 & 344.29  \\ 
Clone             & \bf 50.66 & 955.97 &  46.50  & \bf 666.48 & \bf 48.16 & \bf 1105.11 & 41.28 &  1122.01 & \bf 41.38 & 920.65 & 39.61 & \bf 895.79  \\ 
Authorship        & \bf 91.74 & \bf 447.68   & 73.39 & 1029.59 & \bf 91.17 & \bf 438.69   & 80.39 & 696.64   & \bf 88.88 & \bf 620.56 & 71.79 &  970.44  \\ 
C1000             & \bf 74.70  & \bf 502.02  & 72.96 & 537.76  & \bf 76.82 & \bf 498.64  & 72.23 & 634.27  &  \bf 61.96	& \bf 704.95	   & 59.00    & 697.06  \\ 
Python800         & \bf 83.90  & \bf 460.92  & 77.88 & 514.19 & \bf 79.00    & \bf 496.30  & 71.42 & 730.14  & \bf 72.69	& \bf 646.59  & 69.07 & 662.28  \\ 
Java250           & \bf 79.70  & \bf 760.97  & 75.12 & 815.91 & \bf 81.94 & \bf 744.57  & 72.30  & 853.74  &  \bf 75.52	& \bf 910.97  & 63.80  & 1049.46 \\ \hline
Count & 6/6 & 4/6 & 0/6 & 2/6 & 6/6 & 6/6 & 0/6 & 0/6 & 6/6 & 5/6 & 0/6 & 1/6 \\ \hline
\end{tabular}
}
\caption{Results of ablation study, before and after removing constraints, ASR \%.}
\label{tab:ablation}
\vspace{-2mm}
\end{table*}


\subsection{Attack Defended Model and  Retraining}
\textbf{Attack Defended Model.} We employ \approach and MHM to attack the three defended models provided by ALERT~\cite{naturalattack2022}. These models are prepared by adversarial fine-tuning. Table~\ref{tab:defended_model} presents the results. We can see that RNNS outperforms MHM in two tasks, and MHM is better in one task. This experiment setting actually is not friendly for \approach because ALERT~\cite{naturalattack2022} uses the replacements from pre-trained models which implicitly have the semantic constraint.

\begin{table}[!t]
\centering
\scalebox{0.7}{
\begin{tabular}{l|cc|cc}
\hline
\multirow{2}{*}{Defended Model} & \multicolumn{2}{c|}{RNNS}   & \multicolumn{2}{c}{MHM }      \\ 
                               & ASR & QT  & ASR  & QT \\ \hline
Clone+CodeBert                 & 12.90 & \bf 958.35         & \bf 28.17 & 1245.75	       \\ 
Defect+CodeBert                & \bf 95.37 & \bf  282.20    & 92.23 & 283.66      \\ 
Authorship+CodeBert            & \bf  51.88 & 1524.40       & 43.26  & \bf 1026.08      \\ \hline
\end{tabular}}
\caption{Attack defended models, ASR \%.}
\label{tab:defended_model}
\vspace{-2mm}
\end{table}

\begin{table}[!t]
\centering
\scalebox{0.7}{
\begin{tabular}{l|c|c|c|c}
\hline
           & ACC   & ASR(\approach)  & ASR(MHM)   & ASR(ALERT) \\ \hline
Authorship & 90.62 & 19.81 & \bf 23.58 & 14.28 \\ 
Defect     & 65.14 & \bf 40.46 & 23.69 & 24.53 \\ 
Java250    & 97.63 & 19.67 & 6.65  & \bf 42.91 \\ \hline
\end{tabular}}
\caption{Results of contrastive adversarial retraining, model: CodeBERT.}
\label{tab:adv_train}
\vspace{-2mm}
\end{table}

\noindent
\textbf{Retraining.} We use the adversarial examples from \approach to retrain the victim models of CodeBERT by contrastive adversarial learning.  We use three 3 datasets, Defect, Authorship, and Java250. We generate the adversarial examples on the whole training dataset for them. Table~\ref{tab:adv_train} presents the results, all approaches achieve much lower ASR compared with the previous. \approach adversarial examples can improve the mode robustness through contrastive adversarial retraining. If we compare Defect/Authorship+CodeBERT in \tabref{adv_train} and \tabref{defended_model}, we can find that both retrained models via RNNS are more robust than the models from ALERT since they have much lower ASRs. 

\subsection{RNNS  vs Textual Attack Methods}

\cameraready{To compare the effects of RNNS and textual attack methods, We conducted attack experiments on three datasets using the PSO~\citep{zang2020word} and LSH~\citep{maheshwary2021strong}. The three datasets Defect, Authorship, and Java250, represent three languages respectively, C, Python, and Java. To be fair, the search space of the PSO and LSH is the same as that of RNNS.}

\cameraready{As shown in \tabref{PSO}, the QT of PSO algorithm is 4.22-6.7 times that of RNNS, and the ASR of PSQ algorithm is 5.55\% - 27.82\% lower than that of RNNS algorithm. It can be inferred that for code variable attacks, combinatorial optimization is inefficient when the substitute set of variables is relatively large. The main reasons are the following two points. Firstly, code segments are generally longer, and the substitute set of code variables is much larger than the synonym set of natural language words. Secondly, the impact of variable replacement on code semantics is smaller than that of word replacement on natural language semantics. }

\begin{table}[!t]
\centering
\scalebox{0.6}{
\begin{tabular}{l|cc|cc|cc}
\hline
\multirow{2}{*}{Task+Model} & \multicolumn{2}{c|}{RNNS}   & \multicolumn{2}{c|}{PSO}    & \multicolumn{2}{c}{LSH}  \\ 
                               & ASR & QT  & ASR  & QT  & ASR  & QT \\ \hline
Defect+CodeBert                 & \bf 69.18 & 588.35         &  63.63 & 3945.04         &  26.62 & \bf 321.78	       \\ 
Authorship+CodeBert                & \bf 73.39 & 1029.59     & 52.29 & 4350.00     & 19.26 & \bf 458.55      \\ 
Java250+CodeBert            & \bf  75.12 & 815.91       & 47.3 & 5076.02       & 31.58 & \bf 397.05      \\ \hline
\end{tabular}}
\caption{RNNS vs PSO and LSH, ASR \%.}
\label{tab:PSO}
\vspace{-4mm}
\end{table}

\cameraready{RNNS’s QT is 1.8-2.2 times that of LSH, and the QT has dropped significantly. However, LSH’s ASR is inferior to RNNS by 42.56\%-54.13\%. For code variable attacks, LSH has high efficiency, but its effectiveness is relatively low. One possible reason for LSH causing low ASR is the distribution of adversarial samples in each bucket is uneven.}

\section{Related Work}
\label{related_work}
Adversarial attacks for code models have been widely studied~\citep{naturalattack2022, liu2023contrabert, li2023multi, jha2022codeattack}. These works can be generally categorized into black-box attacks and white-box attacks. A black-box attack for code models queries the model outputs and selects the substitutes using a score function. For example, ALERT~\citep{naturalattack2022} finds the adversarial examples using variable-name substitutes generated by pre-trained masked models. MHM~\citep{zhang2020generating} uses Metropolis–Hastings to sample the replacement of code identifiers. STRATA~\citep{springer2020strata} generates adversarial examples by replacing the code tokens based on the token distribution. \citet{9763729} apply pre-defined semantics-preserving code transformations to attack code models. 
CodeAttack~\citep{jha2022codeattack} uses code structure to generate adversarial data. 
White-box attacks require the code model gradient to modify inputs for adversarial example generation. CARROT~\citep{10.1145/3511887} selects code mutated variants based on the model gradient. \citet{ramakrishnan2020semantic} attack code models by gradient-based optimization of the abstract syntax tree transformation. \citet{srikant2021generating} uses optimized program obfuscations to modify the code. DAMP~\citep{yefet2020adversarial} derives the desired wrong prediction by changing inputs guided by the model gradient.

\tabref{com_rnns} demonstrates the differences among \approach, MHM~\citep{zhang2020generating} and ALERT~\citep{naturalattack2022}. MHM and ALERT represent the two methodologies most closely aligned with our research. Our approach considers identifier replacements like MHM and ALERT, ensuring that the adversarial example keeps the same semantics as the original one. Our substitute size is scalable and can be substantial, and \approach searches the possible next adversarial example in the substitute space. In our approach, we locate vulnerable variables based on the uncertainty and search $subs_{topk}$ without building adversarial samples and actual attacks. Our goal is to obtain high ASRs by searching real variable names. MHM has the same goal as ours but synthesizes variable names. ALERT sacrifices ASR to make the variable name readable.

\begin{table}[!t]
\centering
\scalebox{0.6}{
\begin{tabular}[]{c|c|c|c|c}
\hline
Algorithm  & \begin{tabular}[c]{@{}c@{}}Substitutes \\ Size\end{tabular} & \begin{tabular}[c]{@{}c@{}}Substitutes \\ Source\end{tabular}       & \begin{tabular}[c]{@{}c@{}}Replacement \\ Position\end{tabular} & \begin{tabular}[c]{@{}c@{}}Substitutes \\ Selection\end{tabular}            \\ \hline
MHM        & medium                                                      & vocabulary                                                         & random                                                          & \begin{tabular}[c]{@{}c@{}}random \\ sample\end{tabular}                   \\ 
ALERT      & small                                                       & \begin{tabular}[c]{@{}c@{}}model \\ generation\end{tabular}        & \begin{tabular}[c]{@{}c@{}}importance \\ score\end{tabular}     & traverse                                                                   \\ 
RNNS & large                                                       & \begin{tabular}[c]{@{}c@{}}real \\ public\\ variables\end{tabular} & \begin{tabular}[c]{@{}c@{}}uncertainty \\ score\end{tabular}    & \begin{tabular}[c]{@{}c@{}}efficient \\ constrained \\ search\end{tabular} \\ \hline
\end{tabular}}
\caption{Difference between \approach to the others.}
\label{tab:com_rnns}
\vspace{-4mm}
\end{table}

\section{Conclusion}
\label{sec:conclusions}
 We propose a novel black-box adversarial search-based attack for variable replacement. \approach has three main contributions: 1) This work proposes a non-generation search-based black-box attacking method via predicting the attack effect of a substitute. This method can greatly reduce the verification cost of the substitute, remove the restrictions on the size and diversity of the substitute set, and achieve a significant improvement in terms of ASR without increasing QT. 2) This work proposes a simple and efficient method for constructing a substitute set. This method can construct a large-scale, diverse, and real substitute set at low cost. 3) The adversarial examples from \approach can be used to improve the model robustness. 
 

\section{Limitations}
 There are some limitations of \approach. Firstly, RNNS does not revert to the preceding step to persist with the search upon an increase in the model probability of the ground truth label. While the incorporation of this step may bolster the Attack Success Rate (ASR), it could potentially compromise the Query Time (QT). Secondly, the size and diversity of the substitute set significantly influence RNNS; a minimal and homogeneous set can precipitate a diminished attack success rate. Thirdly, RNNS involves multiple hyperparameters whose values need to be manually set. One of the most important parameters is the moving parameter $\alpha$. The number of attacking iterations $max\_itr$ is also significant. We set $\alpha$ to 0.2 and $max\_itr$  to 6 with some small experimental trials. \cameraready{Fourthly, RNNS currently only targets untargeted attack scenarios, for targeted attacks, ASR will be very low when there are many category labels. For example, when performing targeted attacks on Authorship+Codebert with 66 labels, the ASR can only reach 6.4\%.  How to migrate to targeted attacks is a direction we need to study in the future.} 

\section*{Acknowledgment}
This work is supported by NRF and the CSA under its National Cybersecurity R\&D Programme (NCRP25-P04-TAICeN), NRF and DSO National Laboratories under the AI Singapore Programme (AISG Award No: AISG2-RP-2020-019), and NRF Investigatorship NRF-NRFI06-2020-0001. Any opinions, findings and conclusions or recommendations expressed in this material are those of the author(s) and do not reflect the views of NRF and CSA Singapore.


\bibliography{anthology,custom}

\begin{thebibliography}{28}
\expandafter\ifx\csname natexlab\endcsname\relax\def\natexlab#1{#1}\fi

\bibitem[{Ahmad et~al.(2020)Ahmad, Chakraborty, Ray, and
  Chang}]{ahmad2020transformer}
Wasi Ahmad, Saikat Chakraborty, Baishakhi Ray, and Kai-Wei Chang. 2020.
\newblock A transformer-based approach for source code summarization.
\newblock In \emph{Proceedings of the 58th Annual Meeting of the Association
  for Computational Linguistics}, pages 4998--5007.

\bibitem[{Alsulami et~al.(2017)Alsulami, Dauber, Harang, Mancoridis, and
  Greenstadt}]{10.1007/978-3-319-66402-6_6}
Bander Alsulami, Edwin Dauber, Richard Harang, Spiros Mancoridis, and Rachel
  Greenstadt. 2017.
\newblock Source code authorship attribution using long short-term memory based
  networks.
\newblock In \emph{Computer Security -- ESORICS 2017}, pages 65--82, Cham.
  Springer International Publishing.

\bibitem[{Chen et~al.(2022)Chen, Li, Wen, and Liu}]{9763729}
Penglong Chen, Zhen Li, Yu~Wen, and Lili Liu. 2022.
\newblock \href {https://doi.org/10.1109/ICECCS54210.2022.00029} {Generating
  adversarial source programs using important tokens-based structural
  transformations}.
\newblock In \emph{2022 26th International Conference on Engineering of Complex
  Computer Systems (ICECCS)}, pages 173--182.

\bibitem[{Feng et~al.(2020)Feng, Guo, Tang, Duan, Feng, Gong, Shou, Qin, Liu,
  Jiang et~al.}]{feng2020codebert}
Zhangyin Feng, Daya Guo, Duyu Tang, Nan Duan, Xiaocheng Feng, Ming Gong, Linjun
  Shou, Bing Qin, Ting Liu, Daxin Jiang, et~al. 2020.
\newblock Codebert: A pre-trained model for programming and natural languages.
\newblock In \emph{Findings of the Association for Computational Linguistics:
  EMNLP 2020}, pages 1536--1547.

\bibitem[{Gu et~al.(2018)Gu, Zhang, and Kim}]{gu2018deep}
Xiaodong Gu, Hongyu Zhang, and Sunghun Kim. 2018.
\newblock Deep code search.
\newblock In \emph{2018 IEEE/ACM 40th International Conference on Software
  Engineering (ICSE)}, pages 933--944. IEEE.

\bibitem[{Guo et~al.(2020)Guo, Ren, Lu, Feng, Tang, Shujie, Zhou, Duan,
  Svyatkovskiy, Fu et~al.}]{guo2020graphcodebert}
Daya Guo, Shuo Ren, Shuai Lu, Zhangyin Feng, Duyu Tang, LIU Shujie, Long Zhou,
  Nan Duan, Alexey Svyatkovskiy, Shengyu Fu, et~al. 2020.
\newblock Graphcodebert: Pre-training code representations with data flow.
\newblock In \emph{International Conference on Learning Representations}.

\bibitem[{Henkel et~al.(2022)Henkel, Ramakrishnan, Wang, Albarghouthi, Jha, and
  Reps}]{ramakrishnan2020semantic}
Jordan Henkel, Goutham Ramakrishnan, Zi~Wang, Aws Albarghouthi, Somesh Jha, and
  Thomas Reps. 2022.
\newblock \href {https://doi.org/10.1109/SANER53432.2022.00070} {Semantic
  robustness of models of source code}.
\newblock In \emph{2022 IEEE International Conference on Software Analysis,
  Evolution and Reengineering (SANER)}, pages 526--537.

\bibitem[{Jha and Reddy(2023)}]{jha2022codeattack}
Akshita Jha and Chandan~K Reddy. 2023.
\newblock Codeattack: Code-based adversarial attacks for pre-trained
  programming language models.
\newblock In \emph{Proceedings of the AAAI Conference on Artificial
  Intelligence}, volume~37, pages 14892--14900.

\bibitem[{Li et~al.(2017)Li, Feng, Zhuang, Meng, and Ryder}]{li2017cclearner}
Liuqing Li, He~Feng, Wenjie Zhuang, Na~Meng, and Barbara Ryder. 2017.
\newblock Cclearner: A deep learning-based clone detection approach.
\newblock In \emph{2017 IEEE International Conference on Software Maintenance
  and Evolution (ICSME)}, pages 249--260. IEEE.

\bibitem[{Li et~al.(2023)Li, Liu, Chen, Xie, Zhang, and Liu}]{li2023multi}
Yanzhou Li, Shangqing Liu, Kangjie Chen, Xiaofei Xie, Tianwei Zhang, and Yang
  Liu. 2023.
\newblock \href {https://doi.org/10.18653/v1/2023.acl-long.399} {Multi-target
  backdoor attacks for code pre-trained models}.
\newblock In \emph{Proceedings of the 61st Annual Meeting of the Association
  for Computational Linguistics (Volume 1: Long Papers)}, pages 7236--7254,
  Toronto, Canada. Association for Computational Linguistics.

\bibitem[{Li et~al.(2022)Li, Qi, Gao, Peng, Lo, Xu, and Lyu}]{li2022closer}
Yaoxian Li, Shiyi Qi, Cuiyun Gao, Yun Peng, David Lo, Zenglin Xu, and Michael~R
  Lyu. 2022.
\newblock A closer look into transformer-based code intelligence through code
  transformation: Challenges and opportunities.
\newblock \emph{arXiv preprint arXiv:2207.04285}.

\bibitem[{Liu et~al.(2020)Liu, Chen, Xie, Siow, and Liu}]{liu2020retrieval}
Shangqing Liu, Yu~Chen, Xiaofei Xie, Jing~Kai Siow, and Yang Liu. 2020.
\newblock Retrieval-augmented generation for code summarization via hybrid gnn.
\newblock In \emph{International Conference on Learning Representations}.

\bibitem[{Liu et~al.(2023{\natexlab{a}})Liu, Wu, Xie, Meng, and
  Liu}]{liu2023contrabert}
Shangqing Liu, Bozhi Wu, Xiaofei Xie, Guozhu Meng, and Yang Liu.
  2023{\natexlab{a}}.
\newblock \href {https://doi.org/10.1109/ICSE48619.2023.00207} {Contrabert:
  Enhancing code pre-trained models via contrastive learning}.
\newblock In \emph{2023 IEEE/ACM 45th International Conference on Software
  Engineering (ICSE)}, pages 2476--2487.

\bibitem[{Liu et~al.(2023{\natexlab{b}})Liu, Xie, Siow, Ma, Meng, and
  Liu}]{liu2023graphsearchnet}
Shangqing Liu, Xiaofei Xie, Jingkai Siow, Lei Ma, Guozhu Meng, and Yang Liu.
  2023{\natexlab{b}}.
\newblock Graphsearchnet: Enhancing gnns via capturing global dependencies for
  semantic code search.
\newblock \emph{IEEE Transactions on Software Engineering}.

\bibitem[{Maheshwary et~al.(2021)Maheshwary, Maheshwary, and
  Pudi}]{maheshwary2021strong}
Rishabh Maheshwary, Saket Maheshwary, and Vikram Pudi. 2021.
\newblock A strong baseline for query efficient attacks in a black box setting.
\newblock In \emph{Proceedings of the 2021 Conference on Empirical Methods in
  Natural Language Processing}, pages 8396--8409.

\bibitem[{Puri et~al.(2021)Puri, Kung, Janssen, Zhang, Domeniconi, Zolotov,
  Dolby, Chen, Choudhury, Decker, Thost, Buratti, Pujar, Ramji, Finkler,
  Malaika, and Reiss}]{puri2021codenet}
Ruchir Puri, David Kung, Geert Janssen, Wei Zhang, Giacomo Domeniconi, Vladmir
  Zolotov, Julian Dolby, Jie Chen, Mihir Choudhury, Lindsey Decker, Veronika
  Thost, Luca Buratti, Saurabh Pujar, Shyam Ramji, Ulrich Finkler, Susan
  Malaika, and Frederick Reiss. 2021.
\newblock Codenet: A large-scale ai for code dataset for learning a diversity
  of coding tasks.

\bibitem[{Springer et~al.(2020)Springer, Reinstadler, and
  O'Reilly}]{springer2020strata}
Jacob~M Springer, Bryn~Marie Reinstadler, and Una-May O'Reilly. 2020.
\newblock Strata: Simple, gradient-free attacks for models of code.
\newblock \emph{arXiv preprint arXiv:2009.13562}.

\bibitem[{Srikant et~al.(2021)Srikant, Liu, Mitrovska, Chang, Fan, Zhang, and
  O'Reilly}]{srikant2021generating}
Shashank Srikant, Sijia Liu, Tamara Mitrovska, Shiyu Chang, Quanfu Fan, Gaoyuan
  Zhang, and Una-May O'Reilly. 2021.
\newblock \href {https://openreview.net/forum?id=PH5PH9ZO_4} {Generating
  adversarial computer programs using optimized obfuscations}.
\newblock In \emph{International Conference on Learning Representations}.

\bibitem[{Wang et~al.(2020)Wang, Li, Ma, Xia, and Jin}]{wang2020detecting}
Wenhan Wang, Ge~Li, Bo~Ma, Xin Xia, and Zhi Jin. 2020.
\newblock Detecting code clones with graph neural network and flow-augmented
  abstract syntax tree.
\newblock In \emph{2020 IEEE 27th International Conference on Software
  Analysis, Evolution and Reengineering (SANER)}, pages 261--271. IEEE.

\bibitem[{Wang et~al.(2021)Wang, Wang, Joty, and Hoi}]{wang2021codet5}
Yue Wang, Weishi Wang, Shafiq Joty, and Steven~C.H. Hoi. 2021.
\newblock \href {https://doi.org/10.18653/v1/2021.emnlp-main.685} {{C}ode{T}5:
  Identifier-aware unified pre-trained encoder-decoder models for code
  understanding and generation}.
\newblock In \emph{Proceedings of the 2021 Conference on Empirical Methods in
  Natural Language Processing}, pages 8696--8708, Online and Punta Cana,
  Dominican Republic. Association for Computational Linguistics.

\bibitem[{White et~al.(2016)White, Tufano, Vendome, and
  Poshyvanyk}]{white2016deep}
Martin White, Michele Tufano, Christopher Vendome, and Denys Poshyvanyk. 2016.
\newblock Deep learning code fragments for code clone detection.
\newblock In \emph{2016 31st IEEE/ACM International Conference on Automated
  Software Engineering (ASE)}, pages 87--98. IEEE.

\bibitem[{Yang et~al.(2022)Yang, Shi, He, and Lo}]{naturalattack2022}
Zhou Yang, Jieke Shi, Junda He, and David Lo. 2022.
\newblock \href {https://doi.org/10.1145/3510003.3510146} {Natural attack for
  pre-trained models of code}.
\newblock In \emph{Proceedings of the 44th International Conference on Software
  Engineering}, ICSE '22, page 1482–1493, New York, NY, USA. Association for
  Computing Machinery.

\bibitem[{Yefet et~al.(2020)Yefet, Alon, and Yahav}]{yefet2020adversarial}
Noam Yefet, Uri Alon, and Eran Yahav. 2020.
\newblock Adversarial examples for models of code.
\newblock \emph{Proceedings of the ACM on Programming Languages},
  4(OOPSLA):1--30.

\bibitem[{Zang et~al.(2020)Zang, Qi, Yang, Liu, Zhang, Liu, and
  Sun}]{zang2020word}
Yuan Zang, Fanchao Qi, Chenghao Yang, Zhiyuan Liu, Meng Zhang, Qun Liu, and
  Maosong Sun. 2020.
\newblock Word-level textual adversarial attacking as combinatorial
  optimization.
\newblock In \emph{Proceedings of the 58th Annual Meeting of the Association
  for Computational Linguistics}, pages 6066--6080.

\bibitem[{Zhang et~al.(2022)Zhang, Fu, Li, Ma, Zhao, Yang, Sun, Liu, and
  Jin}]{10.1145/3511887}
Huangzhao Zhang, Zhiyi Fu, Ge~Li, Lei Ma, Zhehao Zhao, Hua’an Yang, Yizhe
  Sun, Yang Liu, and Zhi Jin. 2022.
\newblock \href {https://doi.org/10.1145/3511887} {Towards robustness of deep
  program processing models—detection, estimation, and enhancement}.
\newblock \emph{ACM Trans. Softw. Eng. Methodol.}, 31(3).

\bibitem[{Zhang et~al.(2020)Zhang, Li, Li, Ma, Liu, and
  Jin}]{zhang2020generating}
Huangzhao Zhang, Zhuo Li, Ge~Li, Lei Ma, Yang Liu, and Zhi Jin. 2020.
\newblock Generating adversarial examples for holding robustness of source code
  processing models.
\newblock In \emph{Proceedings of the AAAI Conference on Artificial
  Intelligence}, volume~34, pages 1169--1176.

\bibitem[{Zhelezniak et~al.(2020)Zhelezniak, Savkov, and
  Hammerla}]{zhelezniak-etal-2020-estimating}
Vitalii Zhelezniak, Aleksandar Savkov, and Nils Hammerla. 2020.
\newblock \href {https://doi.org/10.18653/v1/2020.acl-main.741} {Estimating
  mutual information between dense word embeddings}.
\newblock In \emph{Proceedings of the 58th Annual Meeting of the Association
  for Computational Linguistics}, pages 8361--8371, Online. Association for
  Computational Linguistics.

\bibitem[{Zhou et~al.(2019)Zhou, Liu, Siow, Du, and Liu}]{zhou2019devign}
Yaqin Zhou, Shangqing Liu, Jingkai Siow, Xiaoning Du, and Yang Liu. 2019.
\newblock Devign: Effective vulnerability identification by learning
  comprehensive program semantics via graph neural networks.
\newblock \emph{Advances in neural information processing systems}, 32.

\end{thebibliography}
\bibliographystyle{acl_natbib}

\end{document}